\documentclass[aip,preprint]{revtex4-1}
\usepackage{graphicx}
\usepackage{xcolor}
\draft 

\begin{document}


\title{Excitation lifetime extracted from electron-photon (EELS-CL) nanosecond-scale temporal coincidences} 



\author{Nadezda Varkentina}
\affiliation{Universit\'e Paris-Saclay, CNRS, Laboratoire de Physique des Solides, 91405, Orsay, France}

\author{Yves Auad}
\affiliation{Universit\'e Paris-Saclay, CNRS, Laboratoire de Physique des Solides, 91405, Orsay, France}

\author{Steffi Y. Woo}
\affiliation{Universit\'e Paris-Saclay, CNRS, Laboratoire de Physique des Solides, 91405, Orsay, France}

\author{Florian Castioni}
\affiliation{Universit\'e Paris-Saclay, CNRS, Laboratoire de Physique des Solides, 91405, Orsay, France}

\author{Jean-Denis Blazit}
\affiliation{Universit\'e Paris-Saclay, CNRS, Laboratoire de Physique des Solides, 91405, Orsay, France}

\author{Marcel Tenc\'e}
\affiliation{Universit\'e Paris-Saclay, CNRS, Laboratoire de Physique des Solides, 91405, Orsay, France}

\author{Huan-Cheng Chang}
\affiliation{Academia Sinica, Institute of Atomic and Molecular Sciences, Taipei, Taiwan}

\author{Jeson Chen}
\affiliation{Academia Sinica, Institute of Atomic and Molecular Sciences, Taipei, Taiwan}

\author{Kenji Watanabe}
\affiliation{Research Center for Electronic and Optical Materials, National Institute for Materials Science, 1-1 Namiki, Tsukuba 305-0044, Japan}

\author{Takashi Taniguchi}
\affiliation{Research Center for Materials Nanoarchitectonics, National Institute for Materials Science,  1-1 Namiki, Tsukuba 305-0044, Japan}

\author{Mathieu Kociak}
\affiliation{Universit\'e Paris-Saclay, CNRS, Laboratoire de Physique des Solides, 91405, Orsay, France}

\author{Luiz H. G. Tizei}
\email[]{luiz.galvao-tizei@universite-paris-saclay.fr}
\affiliation{Universit\'e Paris-Saclay, CNRS, Laboratoire de Physique des Solides, 91405, Orsay, France}


\date{\today}

\begin{abstract}
    Electron-photon temporal correlations in electron energy loss (EELS) and cathodoluminescence (CL) spectroscopies have recently been used to measure the relative quantum efficiency of materials. This combined spectroscopy, named Cathodoluminescence excitation spectroscopy (CLE), allows the identification of excitation and decay channels which are hidden in average measurements. Here, we demonstrate that CLE can also be used to measure excitations' decay time. In addition, the decay time as a function of the excitation energy is accessed, as the energy for each electron-photon pair is probed. We used two well-known insulating materials to characterize this technique, nanodiamonds with \textit{NV$^0$} defects and h-BN with  \textit{4.1 eV} defects. Both also exhibit marked transition radiations, whose extremely short decay times can be used to characterize the instrumental response function. It is found to be typically 2 ns, in agreement with the expected limit of the EELS detector temporal resolution. The measured lifetimes of \textit{NV$^0$} centers in diamond nanoparticles (20 to 40 ns) and \textit{4.1 eV} defect in h-BN flakes ($<$ 2 ns) match those reported previously.

\end{abstract}

\pacs{}

\maketitle 
The specific pathway a system takes to return to its ground state following an optical excitation reveals details about its internal electronic structure \cite{Hill2015, Roquelet2010, Beha2012, Museur2008, Brunner1992}. A key quantity in this type of spectroscopy is the excitation lifetime (the time for its decay). It is defined as the time, $\tau$, by which a single excitation has probability $1/e$ of having decayed. In typical time-dependent optical spectroscopy experiments, photons are used as the excitation source and the decay trace of the generated photoluminescence provides a measurement of $\tau$. Electrons can also be used for spectroscopy in the optical range \cite{Abajo2010, Losquin2015, Polman2019}. Among the possible available techniques, the two mostly used in the optical range are electron energy loss spectroscopy (EELS) which provides a local measurement of optical extinction \cite{Losquin2015}; and cathodoluminescence (CL) which is equivalent to off-resonance photoluminescence \cite{Mahfoud2013}, which can also be used for lifetime measurements \cite{Merano2005, Meuret2016, Liu2016, Meuret2021, Kim2021}. Experiments with electrons have the added benefit of high spatial resolution, down to the nanometer range \cite{Zagonel2011}, and the possibility of exciting non-optical transitions due to the larger momentum carried by electrons \cite{Abajo2010}. Electron excitation is broadband in energy, which can be a benefit, as easy excitation in the far ultra-violet is possible, and a penalty, as the excitation energy cannot be controlled. This penalty can be mitigated by temporal correlation experiments in which the energy lost by each electron leading to a photon emission is measured \cite{Feist2022, Varkentina2022}, a technique coined as Cathodoluminescence Excitation spectroscopy (CLE)\cite{Varkentina2022} as a reference to its photonic counterpart, photoluminescence excitation spectroscopy.
\begin{figure}
 \includegraphics[width=0.8\textwidth]{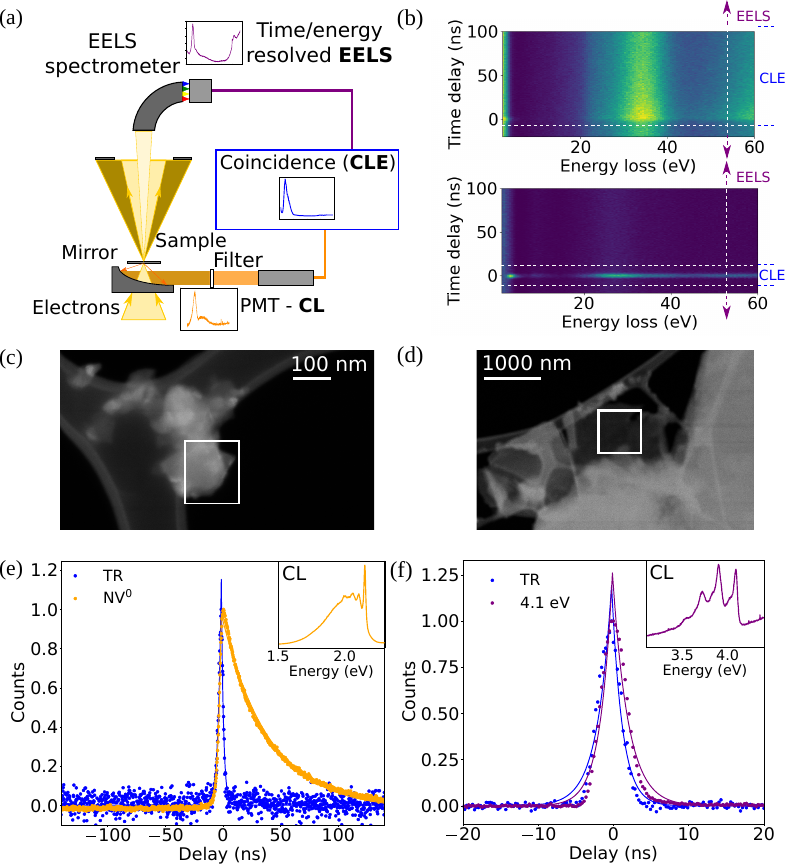}%
 \caption{(a) Sketch of the experiment: an electron microscope is used to produce a 1 nm wide beam of 60 keV electrons. An EELS spectrometer equipped with a magnetic prism measures the transmitted electrons energy spectrum. A parabolic mirror collects light generated by excitations' decay and couples it to a PMT outside the microscope vacuum using a single $3.0 \pm 0.1$ m long multimode 600 $\mu$m-core optical fiber. (b) 2D histograms, $(E; \Delta t)$, for \textit{NV$^0$} in diamond (above) and \textit{4.1 eV} defects in h-BN (below). Because of the shorter lifetimes for the \textit{4.1 eV defects}, coincident electron-photon pairs are less probable for longer time delays in comparison to \textit{NV$^0$} defects. (c-d) Annular dark field images of diamond nanoparticles and an h-BN flake, respectively, supported on an amorphous carbon membrane. (e-f) Temporal traces integrated for all energy losses, $E$, for a diamond nanoparticle and an h-BN flake, with their CL emission spectrum in the inset. The solid lines are a two exponential curve model fitted to the data.  \label{fig::decaycurves}}%
 \end{figure}
 
Concerning only electron spectroscopies, lifetimes have been measured using essentially two methods. The first one uses pulsed electron sources \cite{Merano2005}. In these the electron arrival time on the sample (excitation time) is controlled by the emission time (triggered by a laser \cite{Bostanjoglo2000, Arbouet2018}) or by a fast beam blanker \cite{Bostanjoglo1977, Arbouet2018}. In short, the time delay between a photon emission (CL) and the excitation event is measured, constructing a decay trace \cite{Merano2005}. Lifetime measurements with pulsed electron beams are a more straightforward and efficient way of measuring excitation dynamics. However, this is depreciated by the technical burden of using a pulsed electron gun, well illustrated by the fact that time-resolved CL in a TEM has just recently been demonstrated \cite{Kim2021,Meuret2021}.  In the second method, light intensity interferometry using an Hanbury Brown and Twiss interferometer \cite{Hanbury1993} is used in an electron microscope \cite{Tizei2013} to measure the temporal width of photon bunches emitted by each electron impact, which has been proven to be directly related to the lifetime \cite{Meuret2015}. CL light intensity interferometry has been applied to the measurement of lifetimes at sub-20 nm spatial resolution \cite{Meuret2016} and, in conjunction with CL hyperspectral imaging, to measure the local excitation and emission efficiencies \cite{Meuret2018,Finot2021}. CLE appears as an interesting extension of such approaches, as it could add to the ability to link absorption, emission and dynamics information.

 \begin{figure}
 \includegraphics[width=0.8\textwidth]{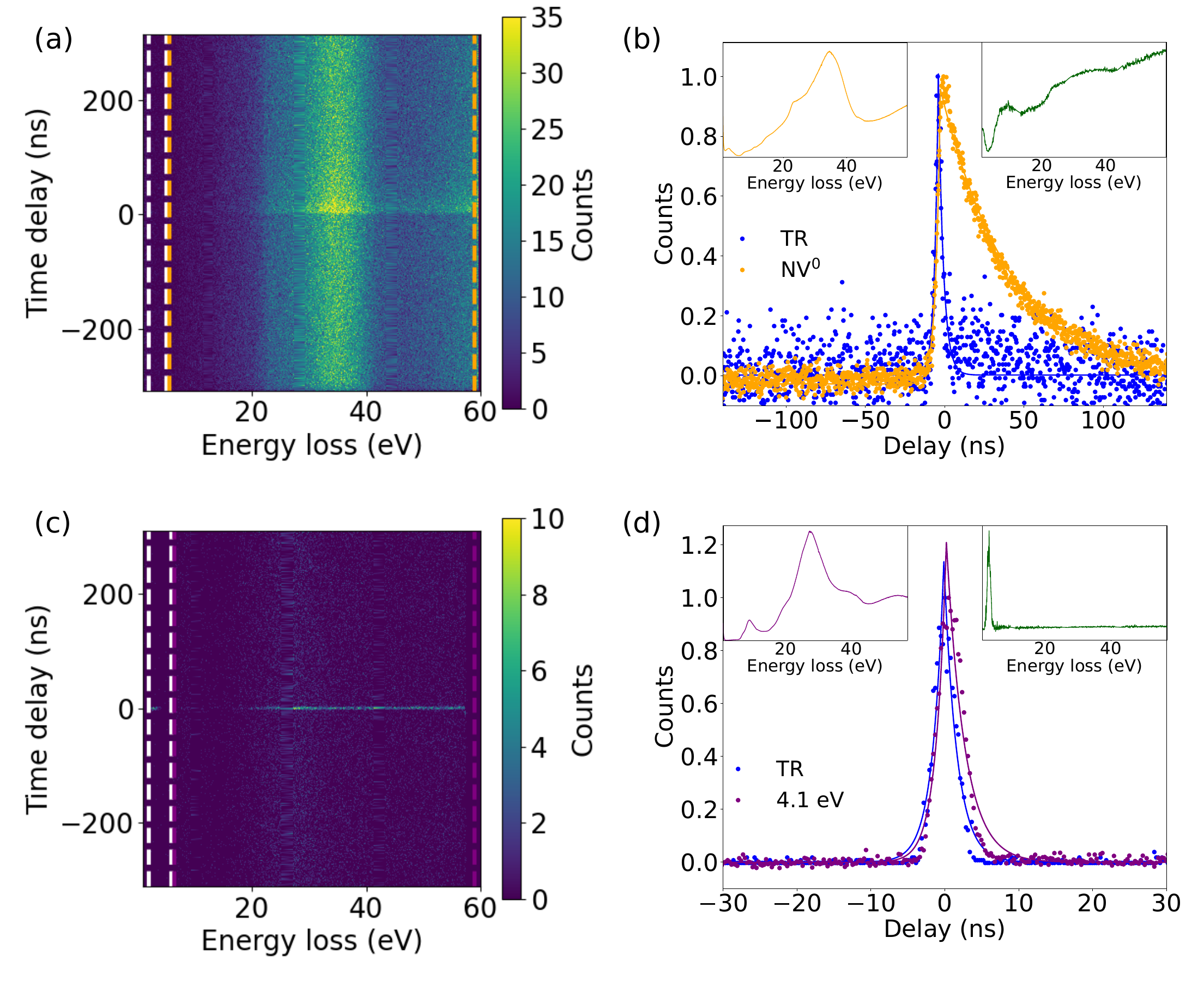}%
 \caption{(a,c) $(E, \Delta t)$ 2D histograms between 1 and 60 eV for \textit{NV$^0$} centers in a diamond nanoparticle and \textit{4.1 eV} defects in an h-BN flake. Colored vertical dashed lines mark energy ranges where temporal profiles are integrated. (b,d) Temporal profiles extracted from (a,c), with the time axis set so decay occurs at positive times. The \textit{NV$^0$} decay is considerably longer than the system IRF. The decay of the \textit{4.1 eV} defects is only marginally longer than the IRF, shown in panel (d) and in Fig. \ref{fig::decaycurves}(f). A diamond nanoparticle and an h-BN flake EELS spectra are shown in the left insets of (b,d). The corresponding relative quantum efficiency curves are shown in the right insets. \label{fig::2DHist}}%
 
 \end{figure}

Here we show that temporal correlations between electron energy loss  (EELS) and photon emission (CL) events provide a measurement of decay traces. We studied two materials: diamond and hexagonal boron nitride (h-BN). To estimate the instrument response function (IRF), we first focused on transition radiation (TR). TR photons are emitted when a fast electron traverses a dielectric interfaces and it occurs at energies in the gap of semiconductors. Its lifetime is known to be much smaller than the measured 2 ns IRF. The decay time of CL emission from \textit{NV$^0$} centers in diamonds nanoparticles was measured to vary between 20 and 40 ns, in agreement with previous measured values using light intensity interferometry \cite{Lourenco2018}. Finally, the decay time of \textit{4.1 eV} defects in h-BN were measured to be barely distinguishable to the IRF, in agreement with previous reports \cite{Meuret2016, Bourrellier2016}.

Experiments were performed on a VG HB501 STEM equipped with a cold field emission electron source, an Attolight M\"onch light injection/collection system and an EELS spectrometer with an ASI Cheetah direct electron detector Fig. \ref{fig::decaycurves}(a). This detector is based on the Timepix3 detector from the Medipix3 consortium, which has recently been used for temporally resolved EELS experiments \cite{Jannis2021, Auad2022, Feist2022, Varkentina2022}. The detector used has 4 chips, aligned in a 4x1 array with 4x(256x256) pixels. In addition, its electronics is equipped with two external time-to-digital-converters. Visible range photons were detected using a photomultiplier tube (PMT H10682-210 single photon counting head from Hamamatsu). The combination of these two time-resolved detectors allows the detection of the time delay between electron scattering and photon detection events as a function of electron energy (Fig. \ref{fig::decaycurves}(b)), as described in detail later. Defects in two materials were studied: NV$^0$ in diamond and \textit{4.1 eV} defects in hexagonal boron nitride (h-BN). In accordance to the emission energy of the corresponding defects, for the diamond experiments a long pass wavelength filter at 532 nm was used, while for the h-BN a short pass wavelength filter at 532 nm was used, unless otherwise noted. 60 keV electrons were focused on a 1 nm-wide probe which was raster scanned on a sample cooled down to 150 K. Electron scattering at high angles ($>$ 80 mrad) allows the formation of high angle annular dark field images, which have an intensity proportional to the projected atomic number, $Z$, and are used for further beam positioning on areas of interest (Fig. \ref{fig::decaycurves}(c-d)). Electron beam convergence and collection angles were 7.5 and 10 mrad. 75-150 nm wide diamond nanoparticles containing a large number ($>$ 100 \textit{NV$^0$} centers) were single crystals produced by proton irradiation of diamond nanoparticles \cite{Yu2005} (Fig. \ref{fig::decaycurves}(c)). h-BN flakes were produced by a high pressure and high temperature method \cite{Taniguchi2007} and chemically exfoliated by sonication in isopropyl alcohol  (Fig. \ref{fig::decaycurves}(d)). The diamond nanoparticles and the thin h-BN flakes were supported on thin amorphous carbon membranes. The CL emission spectrum of these two samples are shown in the insets of Fig. \ref{fig::decaycurves}(e-f) and in the supplementary material (SM) Fig. \ref{figSI::CL_fullrange}. These emissions stem from \textit{NV$^0$} in diamond nanoparticles \cite{Beha2012} and the \textit{4.1 eV} defect in the h-BN flakes \cite{Museur2008}, respectively. In addition to this, a broad emission due to TR occurs \cite{Ginzburg1945}. In the diamond nanoparticles two other minor emissions are observed at 2.5 and 3.2 eV, which are cut by the 532 nm long pass filter (SM Fig. \ref{figSI::CL_fullrange}). For the h-BN flakes, a tail to higher energies was observed, which gives a background to the emission at 4.1 eV (SM Fig. \ref{figSI::CL_fullrange}).

  \begin{figure}
 \includegraphics[width=0.5\textwidth]{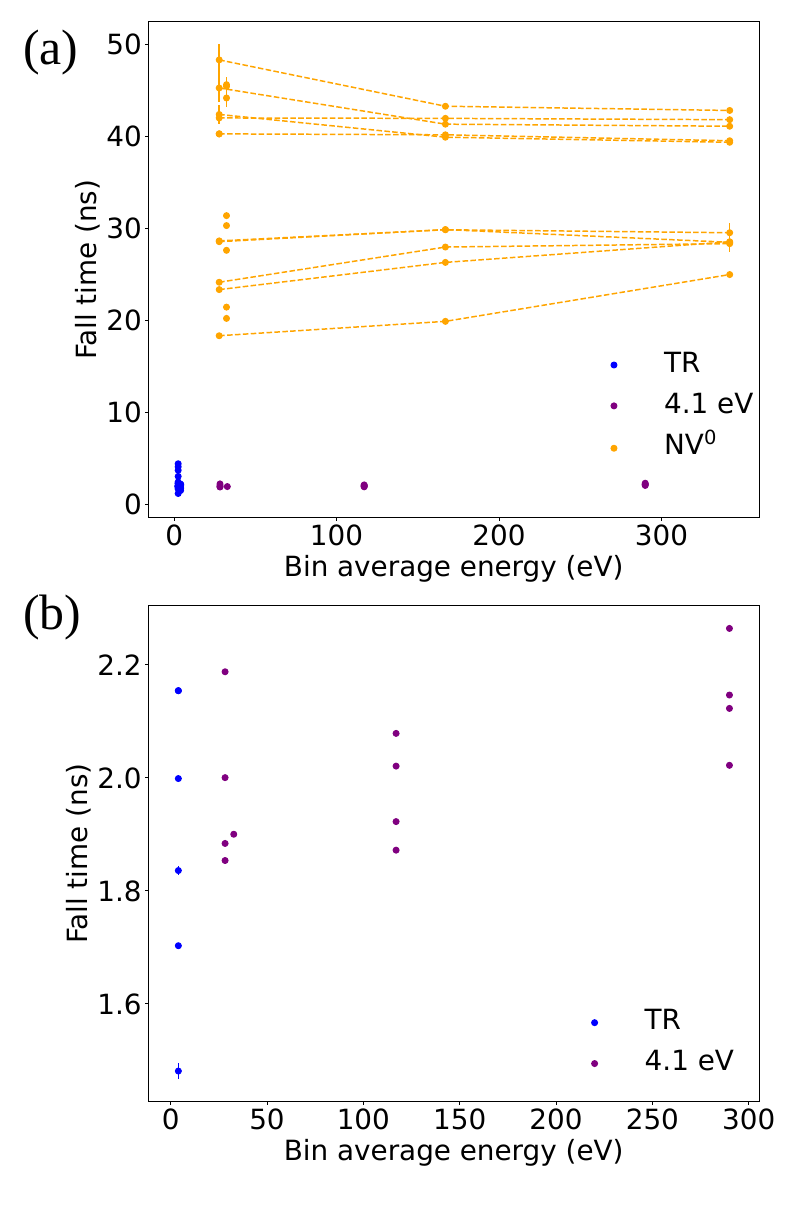}%
 \caption{(a-b) Decay times as a function of energy bins for eighteen \textit{NV$^0$} and five \textit{4.1 eV} defect independent measurements. (b) is the same plot as (a) with only the \textit{4.1 eV} defects in h-BN data. A small but statistically significant variation of the \textit{NV$^0$} decay as a function of excitation energy is observed. Some of the statistical error bars are smaller than the plotted symbols. \label{fig::TimeProfileFullStatistics}}%
 \end{figure}
 
Electrons were detected after a magnetic prism which disperses them along one direction as a function of their energy, with an energy sampling between 0.050 and 1 eV per pixel on the detector. Therefore, for each electron its energy loss after scattering, $E$, and arrival time on the detector, $t_e$, were measured. For photons, only their arrival time, $t_\nu$, was stored. A search algorithm\cite{Varkentina2022} was used to locate temporally correlated events, from which $(E, \Delta t)$ 2D histograms were constructed, with the time delay defined as $\Delta t = t_e - t_{\nu}$. In these histograms (Fig. \ref{fig::decaycurves}(b) and \ref{fig::2DHist}(a,c)) correlated events appear around zero time delay while uncorrelated events appear at longer time delays. These later occur due to various noise sources (detector noise, external particles, such as cosmic rays and ambient photons) and detection losses. For example, if the photon in an electron-photon pair is not detected (for example, if it is emitted away from the collection mirror), the electron which generated it may appear as correlated to a photon from a later scattering event, giving rise to spurious correlations.

An integral of the $(E, \Delta t)$ 2D histograms for all energies results in a temporal profile which represent the rise and decay times of the photon emission probability of an excitation created by an electron inelastic scattering (Figs. \ref{fig::decaycurves}(e-f) and \ref{fig::2DHist}(b-d)).  The time between an electron inelastic event, when an excitation is created, and the first emitted photons may not be zero, leading to a non-zero rise time. For example, a high energy excited state might need to relax, before optical transitions are possible. However, these processes are faster ($<$ 100 ps scale) than our temporal resolution. Following the maximum of the emission intensity, the probability of photon emission will decay, with a typical time scale given by the excitation lifetime. The data presented here was modeled by two exponential functions, from which rise and decay times were extracted. Solid line curves in Fig. \ref{fig::decaycurves}(e-f) are fits using this model. The impulse response function (IRF) is better approximated by a Gaussian curve. However, a model with a Gaussian and two exponential curves have too many free parameters and lead to inconsistent fits. The rise and decay time constant for our instrument response function (IRF) was 2 ns, which was estimated from the temporal profile of the transition radiation (TR). TR photons are emitted when a fast electron traverses a dielectric interfaces and it occurs at energies in the gap of semiconductors. Its lifetime is known to be much smaller than the measured 2 ns IRF \cite{Scheucher2022} (blue curves in Fig. \ref{fig::decaycurves}(e-f)). From the $(E, \Delta t)$ 2D histograms in Fig. \ref{fig::2DHist}(a-c) a CLE spectrum is calculated by summing all electrons leading to photon emission, that is, a projection along the time-delay axis (Fig. \ref{fig::2DHist}(b-d)); total EELS spectrum is the sum of all electrons scattered (at any time); and the relative quantum efficiency (rel. QE \cite{Varkentina2022}) is calculated by dividing the CLE spectrum by the total EELS spectrum, and gives an insight in the preferential electron energy losses responsible for photon emission.  

\begin{figure}
 \includegraphics[width=0.8\textwidth]{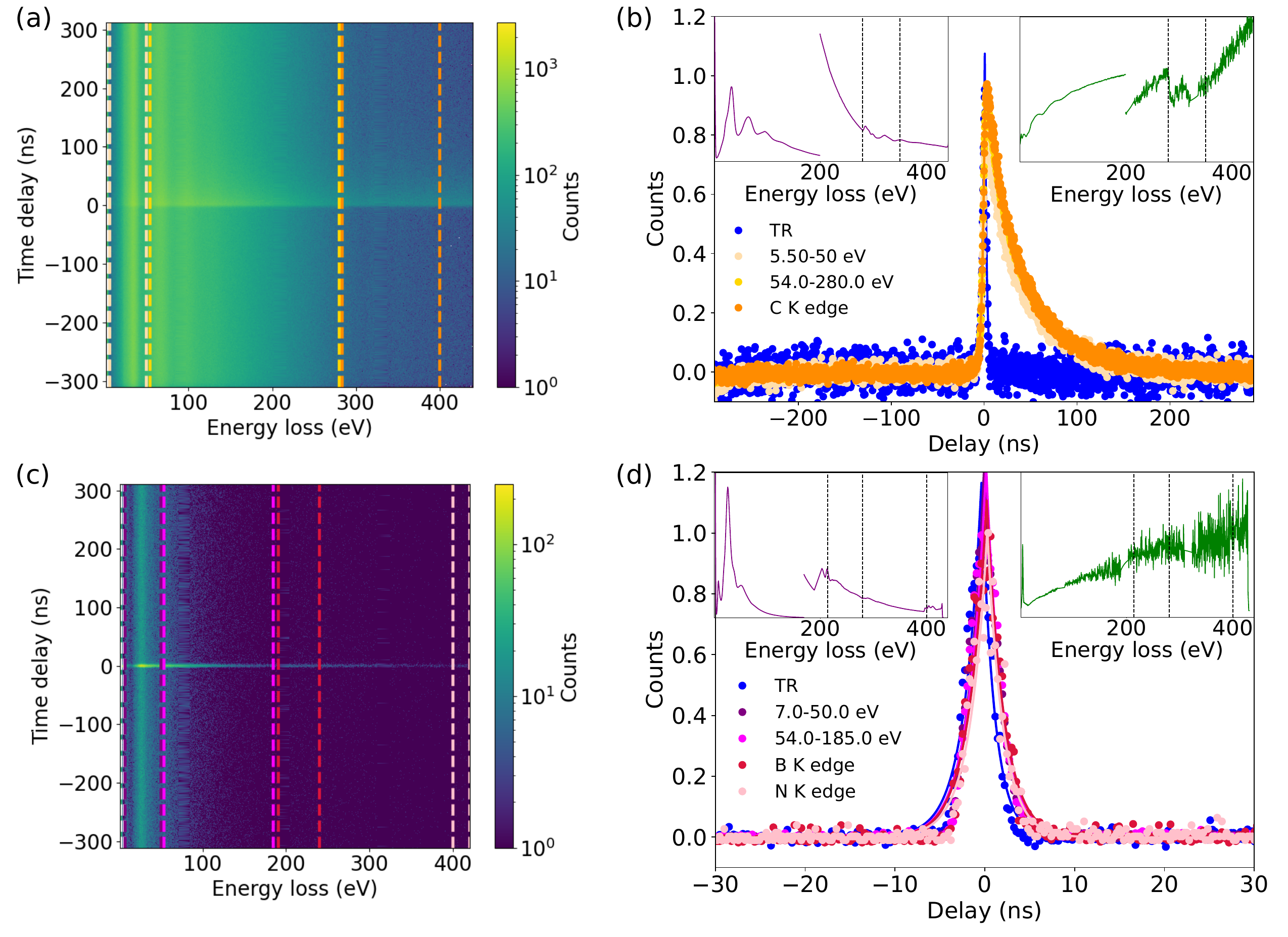}%
 \caption{(a,c)$(E, \Delta t)$ 2D histograms between 1 and ~400 eV for \textit{NV$^0$} centers in a diamond nanoparticle and \textit{4.1 eV} defects in an h-BN flake. Colored vertical dashed lines mark energy ranges where temporal profiles were integrated. (b,d) Temporal profiles extracted from (a,c). A diamond nanoparticle and an h-BN flake EELS spectra are shown in the left insets of (b,d). The colorscale in (a,c) is in a logarithmic scale. The corresponding relative quantum efficiency curves are shown in the right insets.  \label{fig::2DHist_coreloss}}%
 \end{figure}

 
 As a function of electron energy loss (Figs. \ref{fig::2DHist}(a) and \ref{fig::2DHist_coreloss}(a)), two contributions for diamonds are observed: i) between 2.0 and 5.0 eV a fast decay (white vertical lines); and ii) between 6.5 and 440.0 eV a slower decay, orange vertical lines). The corresponding decay profiles are shown in Figs. \ref{fig::2DHist}(b) and \ref{fig::2DHist_coreloss}(a) (with the 2.0 to 5.0 eV decay curve in blue, marked TR).  The fast decay at low electron energy losses is attributed to TR. The slow decay contribution is attributed to the decay of \textit{NV$^0$} centers (20 and 45 ns for the orange profiles in Fig. \ref{fig::2DHist}(c)). In fact, for all energies above the band gap of diamond ($\approx$ 5.5 eV) a similar decay trace is observed. For this reason, the energy integrated temporal profile (Fig. \ref{fig::decaycurves}(e)) has the same long decay. This is expected, as the majority of the emission observed in these diamonds stem from \textit{NV$^0$ centers}. For the h-BN only fast decays were observed (Figs. \ref{fig::2DHist}(c) and \ref{fig::2DHist_coreloss}(c)). Different decay times were observed for \textit{NV$^0$} in diamond nanoparticles (between 20 and 40 ns in Fig. \ref{fig::decaycurves}(e) and \ref{fig::TimeProfileFullStatistics}(a)) and the \textit{4.1 eV defect} in h-BN (2 ns in Fig. \ref{fig::decaycurves}(f)). The data presented in Fig. \ref{fig::TimeProfileFullStatistics} includes eighteen \textit{NV$^0$} and five \textit{4.1 eV} defects independent measurements. Similar decay times were measured when including electron losses up to core-hole excitations (Fig. \ref{fig::2DHist_coreloss}).

The range of lifetimes measured for the \textit{NV$^0$} matches the values measured in CL experiments in the literature \cite{Meuret2016, Lourenco2018, Sola2019}. A similar lifetime range has also been reported in photoluminescence \cite{Storteboom2015}. The reduction in lifetime, which implies an emission rate decrease, was linked to the nanoparticle size, a direct consequence of the presence of an interface close to the emitting dipole \cite{Kreiter2002}. A nanoparticle size dependence has also been observed for NV$^0$ lifetime measured by photoluminescence \cite{Reineck2019, Plakhotnik2018}. For the \textit{4.1 eV} defect in h-BN the measured decay time is barely distinguishable from the IRF of our experiment. However, one can see that their decay traces are slightly longer than that of TR, consistent with the convolution of two decay times of 2 and 1 ns. 

Similar decay traces of the \textit{NV$^0$} centers and the \textit{4.1 eV} defects are observed for losses up to 600 eV, which include core-electron excitation for the carbon, the boron and the nitrogen for h-BN K edges (Fig. \ref{fig::2DHist_coreloss}(a-d). The boron and nitrogen core-electron excitations are not markedly visible in relative quantum efficiency curves (green in upper right inset), despite being visible in the EELS and CLE spectra (Fig. \ref{fig::2DHist_coreloss}(d) upper left insets). The fine structure of the carbon K edge appears in the relative quantum efficiency curve for \textit{NV$^0$} centers in a diamond nanoparticle (Fig. \ref{fig::2DHist_coreloss}(b)). Compared to previously reported data \cite{Varkentina2022}, where no structure was detected in the relative quantum efficiency at core loss edge energies, two improvements were crucial: improved pixel-to-pixel temporal calibration of the TPX3 detector (more details in \cite{Auad2023}) and increased signal-to-noise ratio. The observation of a connection between core-electron excitations and photon emission gives hope to the observation of light emission from individual atoms with atomic resolution. In short, core-electron transitions can be mapped down to the atomic scale \cite{Muller2008, Kimoto2007}, if they occur at sufficiently high energies. Rare earth atoms might be good candidates, given their atomically localized M absorption edges \cite{Tizei2016} and photon emission under electron excitation.

Shifts of the order of 1 or 2 ns of the "zero" time delay exists, which was defined as the peak of the temporal profile for the 2D histogram integrated for all energies. This can be seen in detail in Fig. \ref{fig::2DHist_coreloss}d. These changes are comparable to the temporal sampling and resolution of the current TPX3 implementation for EELS and probably arise due to modal dispersion in the $3.0 \pm 0.1$ m long, 600 $\mu$m core, fiber \cite{Jacomme1975}. With a better temporal resolution and a monomode fiber or free-space detection, one could first confirm if these shifts are physical. More interestingly, with a substantially improved temporal resolution (tens of ps) one could start to investigate the excitation dynamics of materials under electron irradiation and also increase the spatial resolution of CL hyperspectral maps, as already demonstrated for CL using pulsed electron sources \cite{Liu2016}.

To conclude, a method for measuring excitation lifetimes is demonstrated, with a current decay time resolution of around 2 ns. For optimal conditions the IRF can be reduced to 1.6 ns. This is still a factor of 2 smaller than the optimal performance of Timepix3 detectors. The reasons for this discrepancy is discussed in ref. \cite{Auad2023}. In principle, a spatial resolution as good as that of CL experiments \cite{Zagonel2011} is possible \cite{Varkentina2022}. This method can be implemented in electron microscopes equipped with continuous electron sources, as described here, being compatible with current technologies of electron monochromation \cite{Krivanek2014}. The current temporal resolution is limited by the electron detector.  Future upgrades with faster electron detectors (for example, the Timepix4 detector \cite{Heijhoff2022}) will allow experiments in the sub-nanosecond temporal scale.


%

\section*{Supplementary Material}
The supplementary material contain CL spectra of the two defects studied in this manuscript: \textit{NV$^0$} centers in diamond and \textit{4.1 eV defects} in h-BN.

\section*{Acknowledgement}

This project has been funded in part by the National Agency for Research under the program of future investment TEMPOS-CHROMATEM (reference no. ANR-10-EQPX-50) and the JCJC grant SpinE (reference no. ANR-20-CE42-0020) and by the European Union’s Horizon 2020 research and innovation programme under grant agreement No. 823717 (ESTEEM3) and 101017720 (EBEAM). 

\subsection*{Data processing and availability}
All data was processed using the following Python libraries: Numpy 1.23.5, Matplotlib 3.6.2, Scipy 1.10.0, Hyperspy 1.7.3 \cite{francisco_de_la_pena_2020_4294676}. The raw data processing code is available in Zenodo \cite{Auad2022_6346261}.

All published data are available at the Zenodo repository with the following DOI: 10.5281/zenodo.8090907. 

\newpage
\newpage
	\setcounter{page}{1}
	\renewcommand\thefigure{SI\arabic{figure}}
	\setcounter{figure}{0} 
	\setcounter{section}{0}

\section{Supplementary Material}

\subsection{Cathodoluminescence spectra with full energy range}
\begin{figure}[hbtp]
    \centering
    \includegraphics[width=0.8\textwidth]{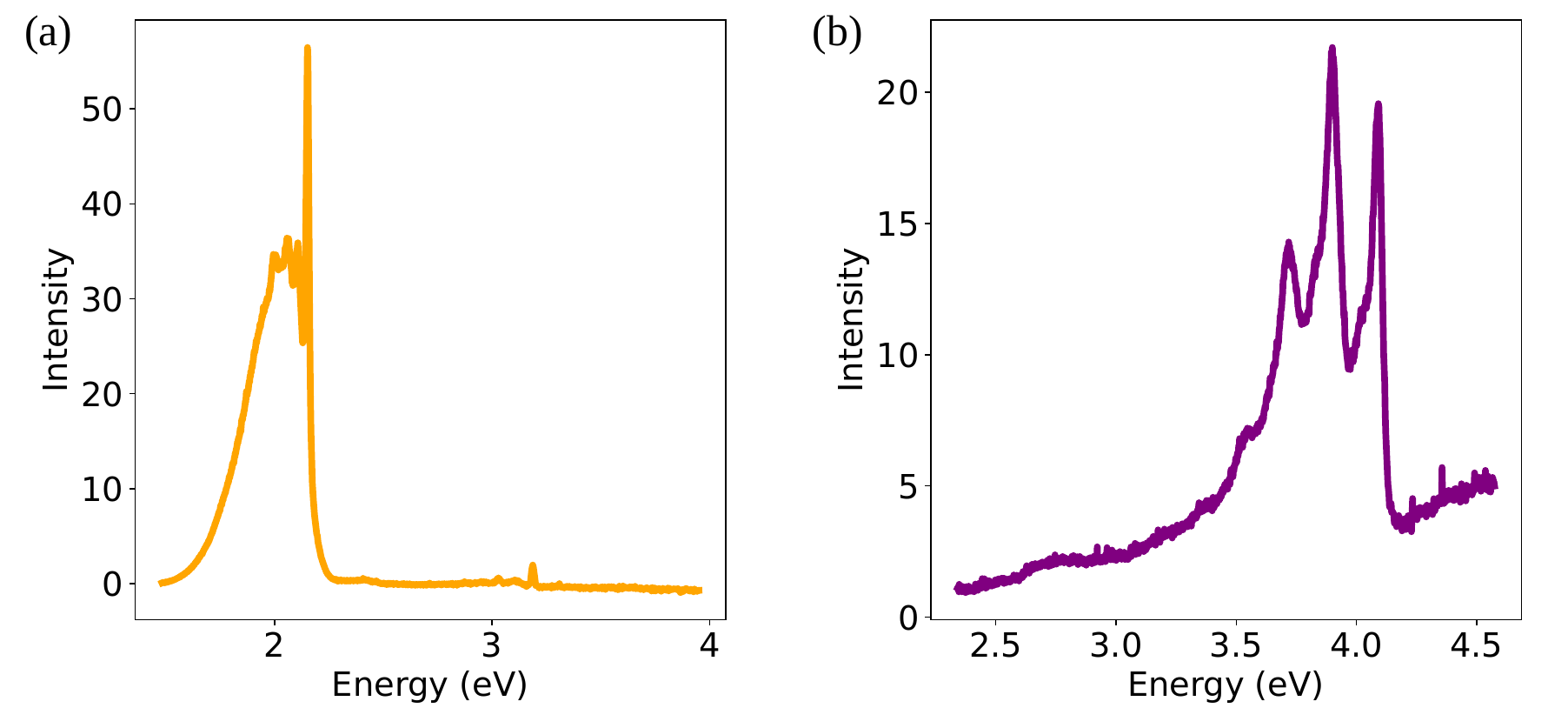}
    \caption{(a-b) CL spectra with their full measured energy range for diamonds containing a large number of NV$^0$ centers and h-BN containing 4.1 eV defects respectively. }
    \label{figSI::CL_fullrange}
\end{figure}

\end{document}